\documentclass[nofootinbib,aps,11pt]{revtex4-1}
\usepackage{graphicx}
\usepackage{hyperref}
\usepackage{cancel}
\usepackage{amssymb}
\usepackage{textcomp}
\usepackage{amsmath}
\usepackage{bm}
\usepackage{times}
\usepackage{epsfig}
\usepackage{color}
\usepackage{subfigure}

\begin{document}
\title{\Large Same-Sign Tetralepton Signature at $\mu$TRISTAN}
\bigskip
\author{Lin-Kun Yan$^1$}
\author{Zhi-Long Han$^1$}
\email{sps\_hanzl@ujn.edu.cn}
\author{Feng-Lan Shao$^2$}
\author{Fa-Xin Yang$^2$}

\affiliation{$^1$School of Physics and Technology, University of Jinan, Jinan, Shandong 250022, China}
\affiliation{$^2$School of Physics and Physical Engineering, Qufu Normal University, Qufu, Shandong 273165, China}
\date{\today}

\begin{abstract}
Naturally tiny neutrino masses can be explained by the low scale seesaw with heavy neutral lepton $N$ coupling to the neutrinophilic Higgs doublet $\Phi_\nu$, which obtains a much smaller vacuum expectation value than the standard Higgs doublet $\Phi$. Within this model, the neutrino masses originate from the new Yukawa interaction $y \overline{L}\tilde{\Phi}_\nu N$. In this paper, we propose the novel same-sign tetralepton signature at the 2 TeV same-sign muon mode $\mu^+\mu^+$ of $\mu$TRISTAN. We investigate two distinct channels of this signature, which are both generated by the Yukawa interaction  $y \overline{L}\tilde{\Phi}_\nu N$. One is from the pair production of charged Higgs $\mu^+\mu^+\to H^+ H^+\to \mu^+N +\mu^+ N\to \mu^+ \mu^+ jj + \mu^+ \mu^+ jj\to 4\mu^+ + 4j$, and the other one is from the single production of charged Higgs $\mu^+\mu^+ \to \mu^+ N H^+ \to \mu^+N +\mu^+ N\to \mu^+ \mu^+ jj + \mu^+ \mu^+ jj\to 4\mu^+ + 4j$. We then perform a detailed simulation of this same-sign tetralepton signature, and obtain the promising region at $\mu$TRISTAN.

\end{abstract}

\maketitle

\section{Introduction}

The observed solar and atmospheric neutrino oscillations indicate that neutrinos have tiny masses below the eV scale \cite{Esteban:2024eli}. However, the origin of such a tiny neutrino mass is still unresolved. The most popular pathway is the well known type-I seesaw mechanism \cite{Minkowski:1977sc,Mohapatra:1979ia,Schechter:1980gr,Schechter:1981cv}, where heavy neutral lepton $N$ is introduced. The Yukawa interaction $y \overline{L} \tilde{\Phi} N$ induces the Dirac mass $m_D= y v / \sqrt{2}$ after the standard Higgs doublet $\Phi$ develops vacuum expectation value (VEV) $\langle \Phi \rangle = v /\sqrt{2}$. Within the canonical seesaw, the light neutrino mass is suppressed by the heavy neutral lepton  as $m_\nu = - v^2 y~ m_N^{-1} y^T /2$. For instance, sub-eV neutrino mass is realized with $y\sim\mathcal{O}(1)$ and $m_N\sim\mathcal{O}(10^{14})$ GeV, but the mass of the heavy neutral lepton $m_N$ is too large to be tested at current or even near future experiments \cite{Abdullahi:2022jlv}.

One promising pathway to lower the scale of heavy neutral lepton is introducing a new Higgs doublet $\Phi_\nu$ with relatively small VEV $v_\nu \ll v$ \cite{Ma:2000cc,Davidson:2009ha,Wang:2016vfj}. Following the original proposal \cite{Ma:2000cc}, the new Higgs doublet $\Phi_\nu$ carries lepton number $L_{\Phi_\nu}=-1$, while the heavy neutral lepton $N$ is assigned with lepton number $L_N=0$. So the canonical Yukawa interaction $y\overline{L} \tilde{\Phi}N$ is forbidden, and the heavy neutral lepton couples to the new Higgs doublet as $y\overline{L} \tilde{\Phi}_\nu N$. After the spontaneous symmetry breaking, the soft $U(1)_L$ violation term $\mu^2(\Phi^\dag \Phi_\nu+\text{h.c.})$ generates a small VEV of $\Phi_\nu$. Light neutrino mass is also generated through the seesaw mechanism as $m_\nu = - v_\nu^2 y~ m_N^{-1} y^T/2$. Typically, sub-eV neutrino mass can be obtained for $y\sim\mathcal{O}(1)$, $v_\nu\sim1$ MeV, and $m_N\sim1$ TeV. With new particles around the TeV scale, various detectable collider signatures of this model have been investigated  \cite{Davidson:2010sf,Haba:2011nb,Seto:2015rma,Guo:2017ybk,Das:2022cmv, Xu:2023hyb,Yang:2024nmk,Okada:2026evy}.

Using the ultra-cold anti-muon technology for muon anomalous magnetic moment at J-PARC \cite{Abe:2019thb}, a 1~TeV $\mu^+$ beam can be realized by re-accelerating. The $\mu^+ e^-$ mode of $\mu$TRISTAN collides at the center-of-mass energy $\sqrt{s}=346$ GeV with $E_\mu=1$ TeV and $E_e=30$ GeV \cite{Hamada:2022mua}. Meanwhile, $\mu$TRISTAN can be also operated in the $\mu^+\mu^+$ mode \cite{Hamada:2022mua}. The $\mu$TRISTAN experiment is promising to test standard model \cite{Hamada:2022uyn,Chen:2024tqh,Hamada:2024ojj,Bhattacharya:2025xwv, Sarkar:2025bgo}  and other new physics \cite{Das:2022mmh,Fukuda:2023yui,Okabe:2023esr, Das:2024gfg, Huang:2025osf,Harigaya:2025zru,Barducci:2025kuq}, such as heavy neutral lepton \cite{Bandyopadhyay:2020mnp,Jiang:2023mte,Santiago:2024zpc,Das:2024kyk,deLima:2024ohf,Dehghani:2025xkd,Kitano:2025xaj,Das:2025rlt}, charged scalar \cite{Fridell:2023gjx,Dev:2023nha,Chiang:2025lab,Li:2025uen,George:2026rzu} , and lepton flavor violation \cite{Yang:2023ojm,Lichtenstein:2023iut,Ding:2024zaj, Calibbi:2024rcm,Kriewald:2024cnt}.

One intriguing signature mediated by the heavy neutral lepton $N$ at the $\mu^+\mu^+$ mode of $\mu$TRISTAN is $\mu^+\mu^+\to W^+W^+$ \cite{Jiang:2023mte,Kitano:2025xaj}, which is the direct evidence of lepton number violation. In the minimal seesaw, the cross section of $\mu^+\mu^+\to W^+W^+$ depends on the mixing parameter $V_{l N}$. The $\mu$TRISTAN experiment is only sensitive to the large mixing region $|V_{l N}|^2\gtrsim10^{-4}$ \cite{Jiang:2023mte}. Therefore, the naturally seesaw predicted value $|V_{l N}|^2\sim m_\nu /m_N\sim10^{-12}$ is far beyond the reach of  $\mu$TRISTAN. In the neutrinophilic Higgs doublet model \cite{Ma:2000cc}, the heavy neutral lepton has additional Yukawa coupling $y\overline{L} \tilde{\Phi}_\nu N$, which could mediate the production of pair charged Higgs $\mu^+\mu^+\to H^+ H^+$ and single charged Higgs $\mu^+\mu^+\to \mu^+ N H^+$ at $\mu$TRISTAN.  Further decays of charged Higgs lead to the same-sign tetralepton signature as $H^+H^+\to \mu^+ N+ \mu^+N\to \mu^+ \mu^+ jj+ \mu^+ \mu^+ jj\to 4\mu^+ +4j$ and $\mu^+ N H^+ \to \mu^+N +\mu^+ N\to \mu^+ \mu^+ jj + \mu^+ \mu^+ jj\to 4\mu^+ + 4j$. In this paper, we investigate the sensitive region of the same-sign tetralepton signature at $\mu$TRISTAN.

This paper is organized as follows. In Section \ref{SEC:MD}, we review the neutrinophilic Higgs doublet model and constraints from lepton flavor violation. The cross sections of the pair production $\mu^+\mu^+\to H^+H^+$ and single production $\mu^+\mu^+\to \mu^+ H^+ N$ are calculated in Section \ref{SEC:CS}. Decay properties of new particles $H^+,N$ and the resulting various collider signatures are considered in \ref{SEC:DP}. The novel same-sign tetralepton signature $4\mu^\pm+4j$ is studied in Section \ref{SEC:SG}. Finally, the conclusion is in Section \ref{SEC:CL}.

\section{The Model}\label{SEC:MD}

In this model, one new Higgs doublet $\Phi_\nu$ with lepton number $L_{\Phi_\nu}=-1$ and three heavy neutral leptons $N$ with $L_N=0$ are introduced. Under the $U(1)_L$ symmetry, the most general scalar potential  is
\begin{eqnarray}\label{Eqn:V}
	V&=-&m_\Phi^2(\Phi^\dag \Phi)+m_{\Phi_\nu}^2(\Phi^\dag_\nu \Phi_\nu)+\frac{1}{2}\lambda_1(\Phi^\dag \Phi)^2+\frac{1}{2}\lambda_2(\Phi^\dag_\nu \Phi_\nu)^2 \\\nonumber 
	&&+\lambda_3(\Phi^\dag \Phi)(\Phi^\dag_\nu \Phi_\nu)+\lambda_4(\Phi^\dag \Phi_\nu)(\Phi^\dag_\nu \Phi)-\mu^2(\Phi^\dag \Phi_\nu+\text{h.c.}),
\end{eqnarray}
where the $\mu^2$-term breaks the $U(1)_L$ symmetry explicitly. The boundedness condition of
the potential requires \cite{Gunion:2002zf}
\begin{equation}
	\lambda_1,\lambda_2>0,~\lambda_3+\sqrt{\lambda_1 \lambda_2}>0,~\lambda_3+\lambda_4+\sqrt{\lambda_1 \lambda_2}>0
\end{equation}
Because the $\mu^2$-term is the only source of lepton number violation, it is naturally small. Such small $\mu^2$-term generates a tiny VEV of neutrinophilic Higgs doublet $\Phi_\nu$ as
\begin{equation}
	v_\nu\simeq \frac{\mu^2 v}{m_{\Phi_\nu}^2+(\lambda_3+\lambda_4)v^2/2}.
\end{equation}
For instance, $v_\nu\sim\mathcal{O}(1)$ MeV can be obtained with $\mu\sim1$ GeV and $m_{\Phi_\nu}\sim 500$ GeV.

After the spontaneous symmetry breaking, the mass eigenstates of scalars are two CP-even Higgs $h$ and $H$, a CP-odd Higgs $A$, and a charged Higgs $H^\pm$.
Masses of these physical Higgs bosons are
\begin{eqnarray}
	m_h^2 &\simeq& \lambda_1 v^2, \label{Eqn:mh}\\  
	m_H^2 &\simeq& m_{\Phi_\nu}^2 + \frac{1}{2}(\lambda_3+\lambda_4)v^2, \label{Eqn:mH}\\  
	m_A^2 &\simeq& m_{\Phi_\nu}^2 + \frac{1}{2}(\lambda_3+\lambda_4)v^2, \\  
	m_{H^\pm}^2 &\simeq& m_{\Phi_\nu}^2 +\frac{1}{2}\lambda_3 v^2,
\end{eqnarray}
where the terms of $\mathcal{O}(v_\nu^2)$ and $\mathcal{O}(\mu^2)$ are neglected. Here, $h$ is regarded as the discovered 125 GeV Higgs at LHC \cite{ATLAS:2015yey}.

In this model, the neutrino sector couples to the new Higgs doublet $\Phi_\nu$ as
\begin{equation}
	-\mathcal{L}_N=y\overline{L}\widetilde{\Phi}_\nu{N}, \label{Eqn:Yukawa}
\end{equation}
with $\widetilde{\Phi}_\nu=i\sigma_2{\Phi}_\nu^*$, which generates the light neutrino mass through the seesaw mechanism
\begin{equation}
	m_\nu=-\frac{1}{2}v_\nu^2{y}m_{N}^{-1}y^T.\label{Eqn:mv}
\end{equation}
By adapting the Casas-Ibarra parametrization \cite{Casas:2001sr}, the Yukawa coupling matrix can be expressed through the neutrino oscillation parameters
\begin{equation}\label{eq:Yuk}
	y = \frac{\sqrt{2}}{v_\nu} U_{\text{PMNS}} \hat{m}_\nu^{1/2} R m_N^{1/2},
\end{equation}
where $U_\text{PMNS}$ is the Pontecorvo-Maki-Nakagawa-Sakata (PMNS) matrix that describing three-neutrino oscillation \cite{Esteban:2024eli}, $\hat{m}_\nu = \text{diag}(m_{\nu 1}, m_{\nu 2}, m_{\nu 3})$, and $R$ is a generalized orthogonal matrix.

With a small value of VEV $v_\nu$, the relatively large Yukawa coupling $y$ not only naturally generates the tiny neutrino mass, but also induces observable charged lepton flavor violation processes \cite{Ma:2001mr,Toma:2013zsa,Bertuzzo:2015ada}. The theoretical branching ratio of $\mu\to e\gamma$ is calculated as \cite{Ding:2014nga}
\begin{equation}
	\text{BR}(\mu\to e\gamma) = \frac{3\alpha}{64\pi G_F^2} \left|\sum_i \frac{y_{\mu i} y_{e i}^*}{m_{H^+}^2} F\left(\frac{m_{N_i}^2}{m_{H^+}^2}\right) \right|^2,
\end{equation}
where $G_F$ is the Fermi constant, and the loop function $F(x)$ is
\begin{equation}
	F(x)=\frac{1}{6(1-x)^4}(1-6x+3x^2+2x^3-6x^2 \ln x).
\end{equation}
Under the current most stringent MEG II limit  BR$(\mu\to e\gamma)<1.5\times 10^{-13}$ (90\% C.L.) \cite{MEGII:2025gzr}, the Yukawa coupling should satisfy
\begin{equation}
	\left|\sum_i y_{\mu i} y_{e i}^* \right|\lesssim 2.6\times10^{-5} \left(\frac{m_{H^+}}{100~\text{GeV}}\right)^2.
\end{equation}
As we consider the muon collider in this paper, we assume $|y_{e i}|\ll |y_{\mu i}|$ to satisfy the lepton flavor violation constraint, which can be obtained by modifying the matrix $R$ via Equation \eqref{eq:Yuk} \cite{Vicente:2014wga,Liu:2022byu}.

\section{Production of Charged Higgs at $\mu$TRISTAN}\label{SEC:CS}

\begin{figure}
	\begin{center}
	\subfigure[]{\includegraphics[width=0.33\linewidth]{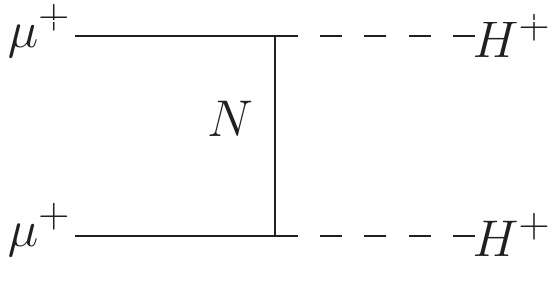}}
	\subfigure[]{\includegraphics[width=0.33\linewidth]{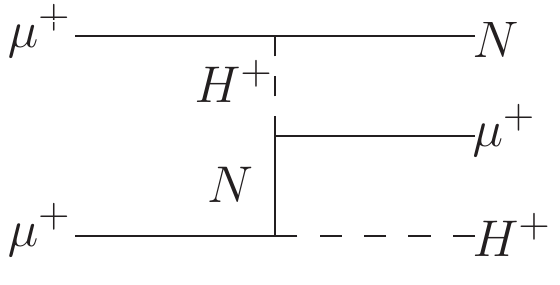}}
	\subfigure[]{\includegraphics[width=0.33\linewidth]{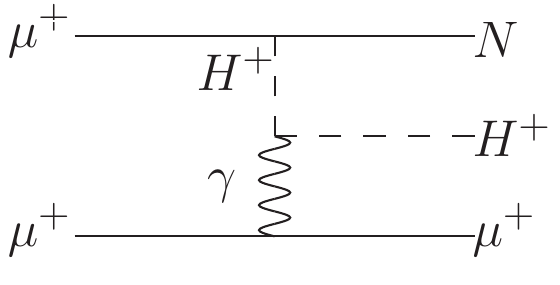}}
	\end{center}
	\caption{Representative Feynman diagrams for pair (panel (a)) and single production (panel (b) and (c)) of charged Higgs at $\mu$TRISTAN.}
	\label{fig1}
\end{figure}

Production of the charged Higgs $H^+$ at $\mu$TRISTAN is dominantly induced by the relatively large muon-flavor Yukawa coupling $y_{\mu N}$.
In Figure \ref{fig1}, we show some representative Feynman diagrams for the pair and single production of charged Higgs. When kinematically allowed, i.e., $2m_{H^+}<\sqrt{s}=2$ TeV, the charge Higgs can be pair produced
\begin{equation}
	\mu^+ \mu^+ \to H^+ H^+ .
\end{equation}
Panel (a) of Figure \ref{fig1} is the $t$-channel production of $H^+H^+$ via heavy neutral lepton $N$, while there is also the $u$-channel contribution. On the other hand, we also have the single production channel
\begin{equation}
	\mu^+ \mu^+ \to \mu^+ H^+ N,
\end{equation}
which is still viable above the pair production threshold $2 m_{H^+}>\sqrt{s}$ when $m_N<m_{H^+}$. Panel (b) of Figure \ref{fig1} is the single production channel via purely Yukawa portal interaction, while panel (c) of Figure \ref{fig1} also involves the gauge portal interaction.

\begin{figure}
	\begin{center}
		\includegraphics[width=0.45\linewidth]{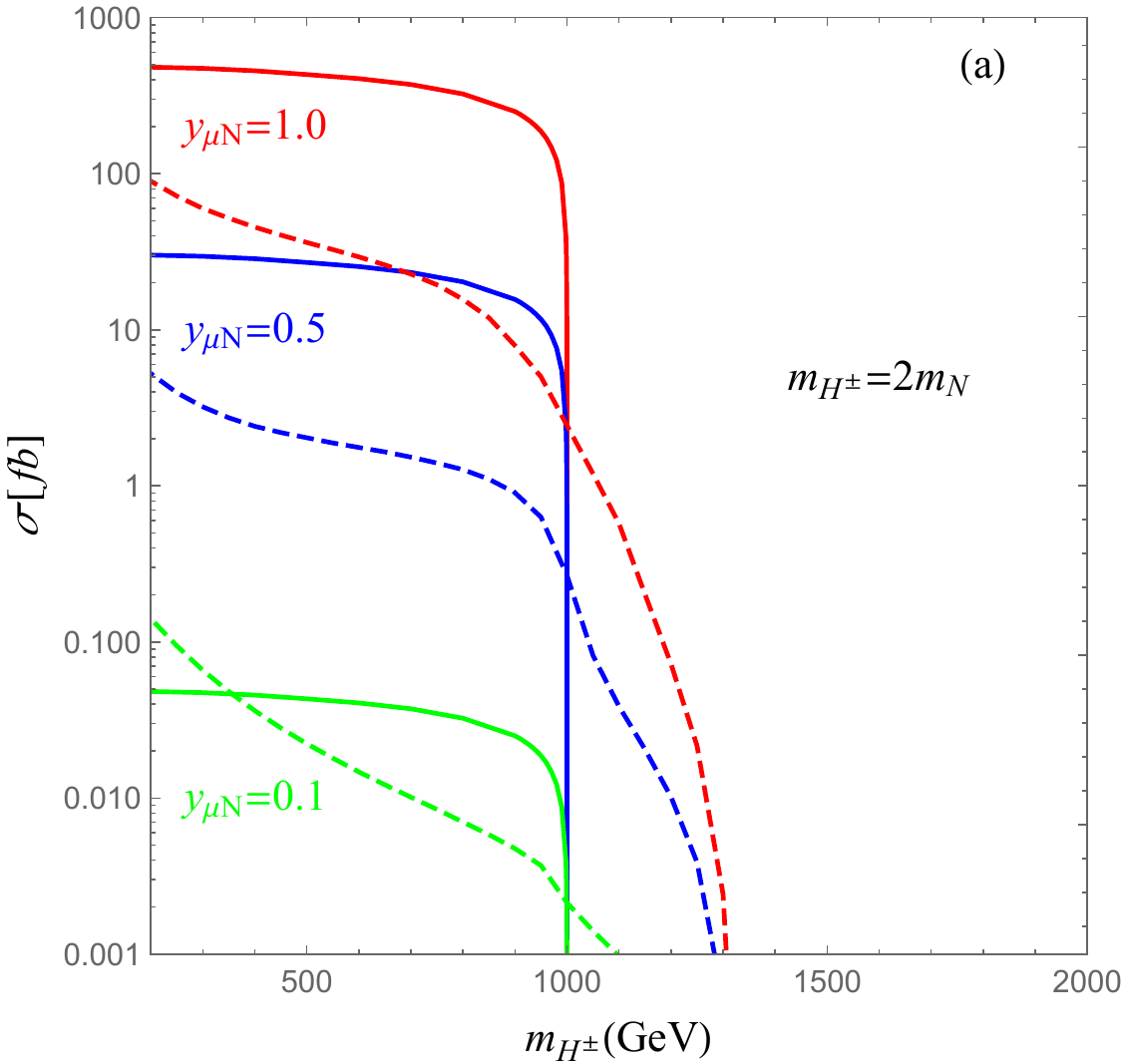}
		\includegraphics[width=0.45\linewidth]{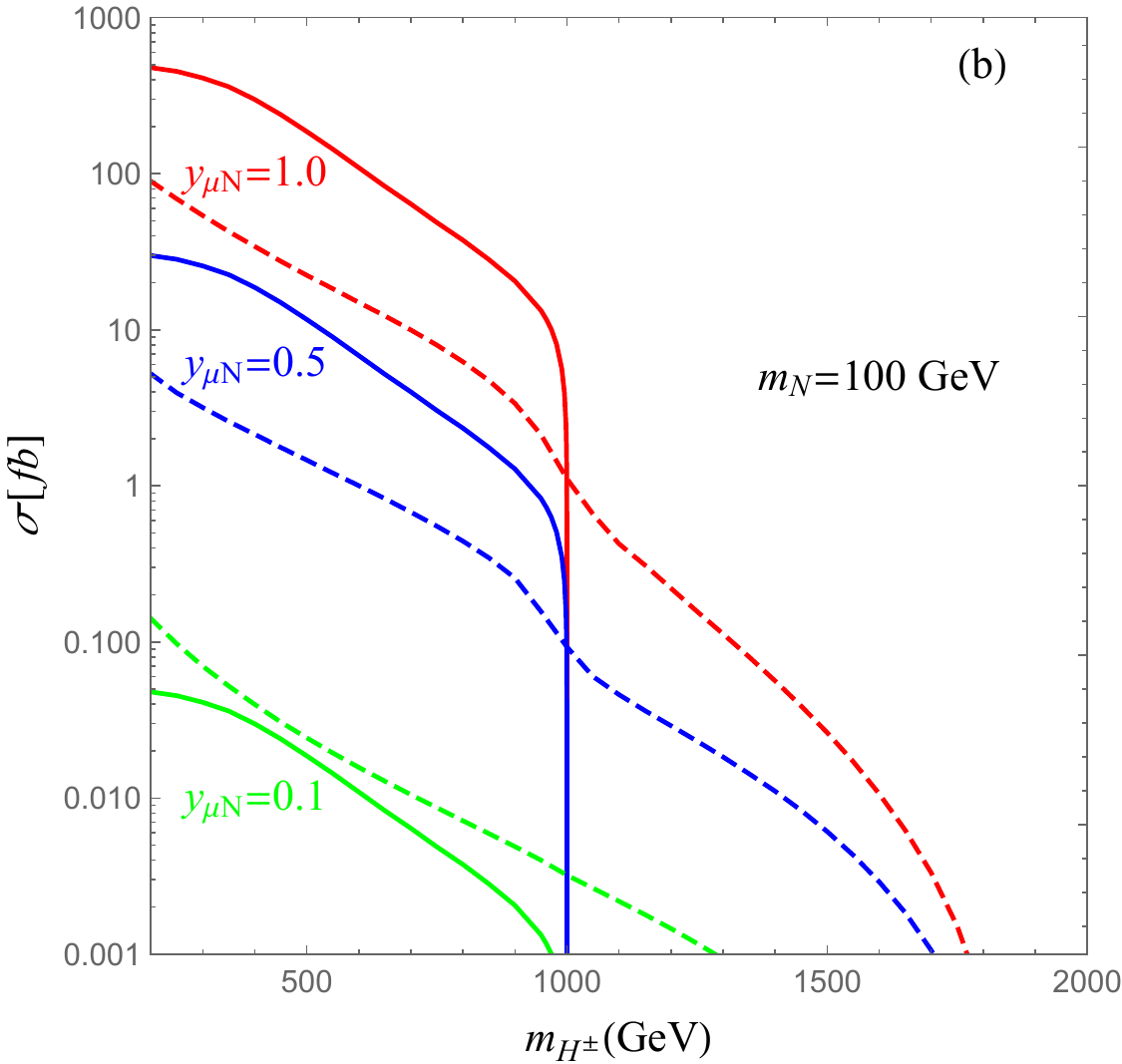}
	\end{center}
	\caption{Cross section of pair and single production of charged Higgs at $\mu$TRISTAN. The solid and dashed lines are the results of pair $\mu^+\mu^+\to H^+ H^+$ and single channel $\mu^+\mu^+ \to \mu^+ H^+ N$, respectively.}
	\label{fig2}
\end{figure}

In Figure \ref{fig2}, we show the cross section of pair and single production of charged Higgs at $\mu$TRISTAN. The related free parameters of these processes are $y_{\mu N}$, $m_N$ and $m_{H^+}$. In case (a) of Figure \ref{fig2}, we fix the mass relation $m_{H^+}=2m_N$. For $y_{\mu N}=1$, the cross section of pair production $\mu^+\mu^+\to H^+ H^+$ is about 450 fb, which gradually decreases when $m_{H^+}$ increases. Suppressed by the three-body final states phase space, the cross section of single production $\mu^+\mu^+\to \mu^+ H^+ N$ is roughly an order of magnitude smaller than pair production when $y_{\mu N}\gtrsim 0.5$. Above the pair production threshold $2 m_{H^+}>\sqrt{s}$, the single production channel still has a sizable cross section. For example, with $y_{\mu N}\gtrsim 0.5$ and $m_{H^+}=2 m_N$, $\sigma(\mu^+\mu^+\to \mu^+ H^+ N)$ is larger than 0.01 fb when $m_{H^+}$ is smaller than about 1200 GeV.

The mass of the heavy neutral lepton also has a significant impact on the production of charged Higgs. In panel (b) of Figure \ref{fig2}, we fix $m_N=100$ GeV. Compared to case (a) with $m_{H^+}=2 m_N$, the cross section of pair production decreases more quickly when $m_{H^+}$ increases. With a larger final state phase space, the cross section of single production decreases more tardily when the charged Higgs is heavier. For instance, $\sigma(\mu^+\mu^+\to \mu^+ H^+ N)$ is about $10^{-3}$ fb when $y_{\mu N}=1$ and $m_{H^+}=1750$ GeV. Therefore, $\mu$TRISTAN has large potential to probe charge Higgs beyond the TeV scale. It is notable that the cross section of pair production $\mu^+\mu^+\to H^+H^+$ is proportional to $|y_{\mu N}|^4$, so as the pure Yukawa portal single production process in panel (b) of Figure \ref{fig1}. For the single production via $\mu^+ \gamma$ collision, the corresponding cross section is only proportional to $|y_{\mu N}|^2$, which is not suppressed heavily for relatively small $y_{\mu N}$. In this way, the single production $\mu^+\mu^+\to \mu^+ H^+ N$ becomes the dominant channel when $y_{\mu N}\lesssim 0.1$.

\section{Decay Property and Various Signatures}\label{SEC:DP}

Decay properties of the new particles heavily depend on the mass spectrum, which then leads to various interesting signatures via cascade decays. First, we consider the
scenario with $m_{H^+}<m_N$, where the Yukawa coupling $y$ mediates the decays of heavy neutral lepton as $N\to \ell^\pm H^\pm,\nu H,\nu A$. Due to mixing between the light and heavy neutrinos, the neutrinophilic scalars could decay as $H^\pm\to \ell^\pm \nu, H/A\to \nu\nu$ when $v/v_\nu\gtrsim \mathcal{O}(10^{5})$ \cite{Haba:2011nb}. Then, pair production of charged Higgs at LHC could generate the dilepton signature $pp\to H^+ H^-\to \ell^+ \nu + \ell^- \nu\to \ell^+\ell^-+\cancel{E}_T$. Direct searches of this dilepton signature have excluded the region with $m_{H^+}\lesssim700$ GeV \cite{ATLAS:2019lff,CMS:2020bfa}. On the other hand, when $v/v_\nu \lesssim \mathcal{O}(10^5)$, the neutrinophilic scalars will decay into SM final states as $H^+\to t\bar{b}, H\to W^+W^-, A\to Zh$ \cite{Batra:2023mds}. Although the LHC has performed the search of charged Higgs via the channel $pp\to t b H^+$ \cite{ATLAS:2018ntn}, the corresponding constraint can be easily satisfied as long as $v_\chi \ll v$ \cite{Batra:2023mds}.

Focus on the muon flavor final states, the pair production of charged Higgs at $\mu$TRISTAN would lead to the same-sign dilepton signature when $v/v_\nu\gtrsim \mathcal{O}(10^{5})$
\begin{equation}
	\mu^+\mu^+\to H^+ H^+ \to \mu^+ \nu +\mu^+\nu.
\end{equation}
This signal is similar as $\mu^+\mu^+\to W^+W^+\to \mu^+ \nu + \mu^+ \nu$, but it can be easily distinguished by the $m_{T2}$ variable \cite{Lester:1999tx}.
Meanwhile, the single production of charged Higgs at $\mu$TRISTAN would induce the tetralepton signature
\begin{equation}
	\mu^+ \mu^+ \to \mu^+ H^+ N \to \mu^+ H^+ + \mu^\pm H^\mp \to \mu^+ \mu^+ \nu + \mu^\pm \mu^\mp \nu.
\end{equation}

When $v/v_\nu\lesssim \mathcal{O}(10^{5})$, the pair and single production of charged Higgs then generate the signature as
\begin{eqnarray}
	\mu^+\mu^+&\to& H^+H^+ \to t\bar{b} + t\bar{b}, \\
	\mu^+\mu^+  \to  \mu^+ H^+ N &\to& \mu^+ H^+ + \mu^\pm H^\mp \to \mu^+ t\bar{b} + \mu^\pm tb.
\end{eqnarray}
Obviously, the pair and single production channels lead to different signatures in this scenario with $m_{H^+}<m_N$. Therefore, these two different channels can be easily distinguished at $\mu$TRISTAN.

For the opposite scenario with $m_{H^+}>m_N$, the decay $H^+\to \ell^+ N$ becomes the dominant channel with relatively large Yukawa coupling, since the couplings of charged Higgs to SM particles are suppressed by the small mixing $v_\nu/v$ \cite{Haba:2011nb}. Then, the heavy neutral lepton decays as $N\to \ell^\pm W^\mp,\nu Z,\nu h$. Production of neutrinophilic scalars at LHC would lead to interesting lepton number violation signatures \cite{Guo:2017ybk}. Since no direct search of such a scenario, we assume $m_{H^+}>200$ GeV to respect the LEP limit \cite{ALEPH:2013htx}.

It is notable that both pair and single production channels could lead to the same signature. For instance, when considering the muon flavor final states, we have
\begin{eqnarray}
	\mu^+ \mu^+ &\to& H^+ H^+ \to \mu^+ N+\mu^+ N, \\
	\mu^+ \mu^+ &\to& \mu^+ H^+ N\to \mu^+ +\mu^+ N+ N.
\end{eqnarray}
 
Further decay of heavy neutral lepton, such as $N\to \mu^\pm W^\mp \to \mu^\pm jj$, generates various interesting signatures
\begin{eqnarray}
	\mu^+\mu^+ NN & \to & \mu^+ \mu^+ + \mu^+ jj +  \mu^+ jj \to 4\mu^+ + 4 j, \\
	& \to & \mu^+ \mu^+ + \mu^- jj + \mu^- jj \to 2\mu^+ + 2 \mu^- + 4j, \\
	& \to & \mu^+ \mu^+ + \mu^+ jj + \mu^- jj \to 3\mu^+ + \mu^- + 4j.
\end{eqnarray}
Here, the same-sign tetralepton $4\mu^++4j$ and opposite-sign tetralepton $2\mu^+ + 2\mu^- +4j$ signatures have the same production cross section.

\section{Same-Sign Tetralepton Signature}\label{SEC:SG}

Both the pair and single production of charged Higgs could lead to the same-sign tetralepton signature at $\mu$TRISTAN. The explicit signal processes are
\begin{eqnarray}
	\mu^+ \mu^+ &\to& H^+ H^+ \to \mu^+ N + \mu^+ N \to \mu^+ \mu^+ jj + \mu^+ \mu^+ jj \\
	\mu^+ \mu^+ &\to& \mu^+ H^+ N \to \mu^+ +\mu^+ N  + N \to \mu^+ + \mu^+ \mu^+ jj +  \mu^+ jj
\end{eqnarray}
There is no irreducible SM background of this signature. Certain contributions might originate from the lepton charge mis-identification. Provided the charge mis-identification rate of 0.1\% \cite{ATLAS:2019jvq}, the $\mu^-$-mistagged cross section of tetralepton signature $\mu^+\mu^+\to \mu^+ \mu^+\mu^+\mu^-$ is about $1.6\times10^{-3}$ fb. Therefore, the SM background is less than $10^{-3}$ fb when additional jets are required in the final states. In the following study, we assume one background event $N_B=1$ for illustration.

In principle, we can distinguish the origins of the same-sign tetralepton signature by fully reconstructing the charged Higgs. However, as already shown in Figure \ref{fig2}, the cross section of single production of charged Higgs is typically much smaller than that of pair production when $y_{\mu N}\sim \mathcal{O}(1)$. Therefore, the same-sign tetralepton signature is dominant by pair production of charged Higgs when $2m_{H^+}<\sqrt{s}$. On the other hand, single production of charged Higgs is the only viable channel above the threshold $2m_{H^+}>\sqrt{s}$. Without the requirement of fully reconstruction of the charged Higgs, we consider the inclusive same-sign tetralepton signature $4\mu^++n j(n\geq2)$ in the following analysis.

We simulate the leading order signal events with {\bf Madgraph5\_aMC@NLO} \cite{Alwall:2014hca} at the parton level. Then, {\bf Pythia8} \cite{Sjostrand:2014zea} is used to do parton showering and hadronization. The detector simulation is performed by {\bf Delphes3} \cite{deFavereau:2013fsa} with the corresponding muon collider card.

With quite a clean SM background, we only apply the following basic cuts:
\begin{eqnarray}
	P_T(\mu^+)>20~\text{GeV},~|\eta(\mu^+)|<2.5, P_T(j)>20~\text{GeV},~|\eta(j)|<2.5.
\end{eqnarray}
The inclusive same-sign tetralepton signature is then selected as
\begin{equation}\label{Eqn:cut}
	N(\mu^+)=4,~N(j)\geq2.
\end{equation}

In Table \ref{Tab01}, we show the results for two benchmark points at the muon collider after the selection cuts in Equation \eqref{Eqn:cut}.
The significance is calculated as~\cite{Cowan:2010js}
\begin{align}
	S=\sqrt{2\left[(N_S+N_B)\ln\left(1+\frac{N_S}{N_B}\right)-N_S\right]},
\end{align}
where $N_S$ and $N_B$ are the events number of the signal and background, respectively. 

\begin{table}
	\begin{center}
		\begin{tabular}{ c | c | c |c} 
			\hline
			\hline
			$\mu^+\mu^+\to 4\mu^++\geq 2j$ &  Cross Section (fb) & Significance & $5\sigma$ Luminosity (fb$^{-1}$)\\
			\hline
			$m_{H^+}=600$ GeV &  213.8 & 619.3 & 0.039 \\
			\hline
			$m_{H^+}=1100$ GeV & 0.17 & 8.37 & 49.0\\
			\hline
			\hline
		\end{tabular}
	\end{center}
	\caption{ Results for the benchmark points at $\mu$TRISTAN, where we have fixed the relation $m_{H^+}=2 m_N$ and $y_{\mu N}=1$. The significance is calculated with an integrated luminosity of 100 fb$^{-1}$. }
	\label{Tab01}
\end{table}

The benchmark of $m_{H^+}=600$ GeV stands for the pair production dominant process. With the Yukawa coupling $y_{\mu N}=1$, the cross section of same-sign tetralepton signature is about 213.8 fb when $m_{H^+}=600$ GeV, which results in a significance of 619.3. Therefore, the pair production channel of same-sign tetralepton signature is quite promising with the Yukawa coupling $y_{\mu N}\sim\mathcal{O}(1)$. Actually, the benchmark of $m_{H^+}=600$ GeV can be discovered with a luminosity of 0.039 fb$^{-1}$.

The benchmark for the single production dominant channel is selected as $m_{H^+}=1100$ GeV. The cross section of the same-sign tetralepton signal for $m_{H^+}=1100$ GeV is about 0.17 fb, which is over three orders of magnitudes smaller than it for $m_{H^+}=600$ GeV. With 100 fb$^{-1}$ data, the significance of $m_{H^+}=1100$~GeV could reach 8.37. And a luminosity of 49.0 fb$^{-1}$ is enough to discover this benchmark point.

\begin{figure}
	\begin{center}
		\includegraphics[width=0.45\linewidth,height=0.45\linewidth]{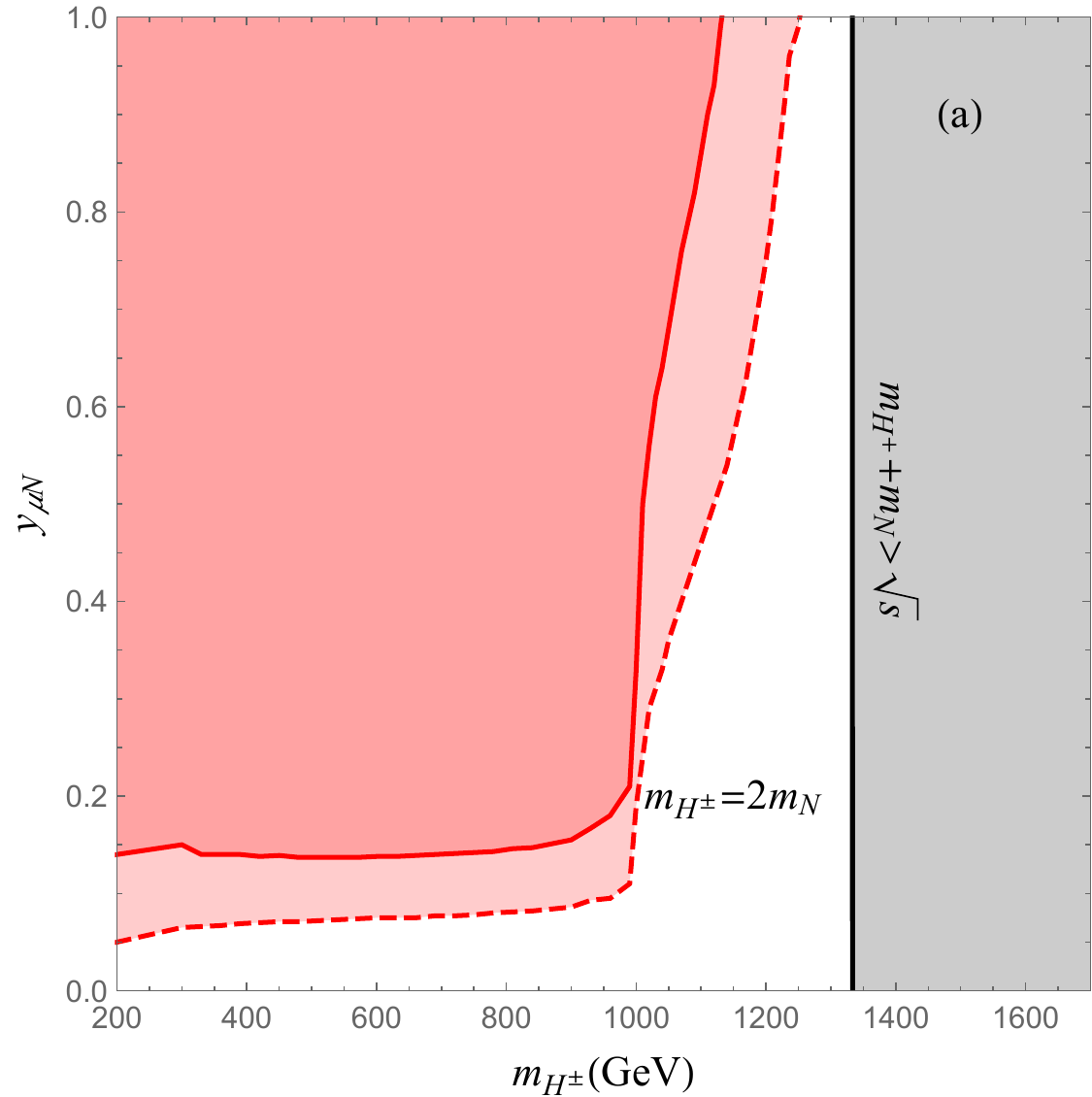}
		\includegraphics[width=0.45\linewidth,height=0.45\linewidth]{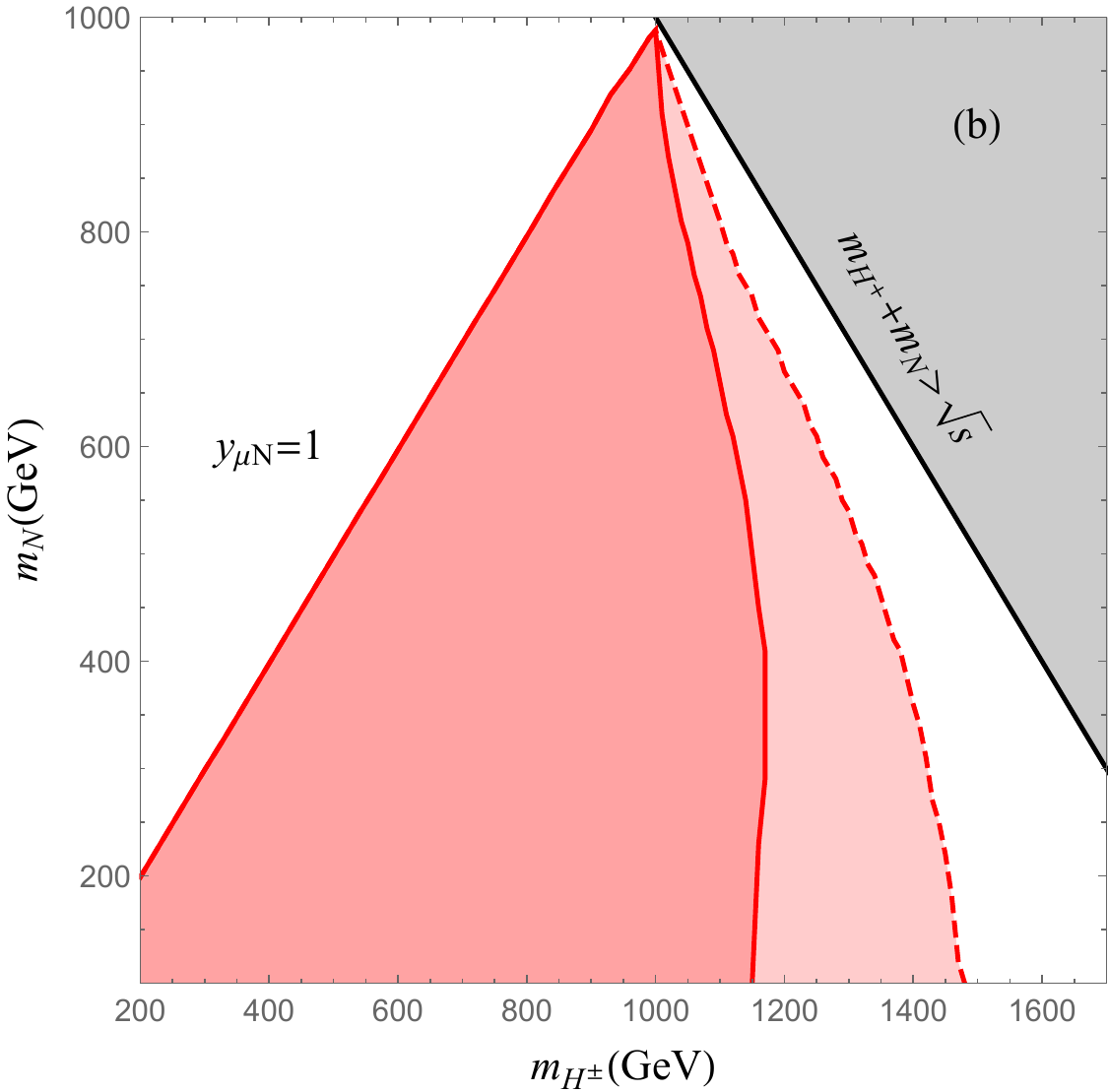}
	\end{center}
	\caption{The $5\sigma$ discovery region of the same-sign tetralepton signature at $\mu$TRISTAN. In panel (a), we fix the mass relation $m_{H^\pm}=2m_N$. In panel (b), the Yukawa coupling $y_{\mu N}=1$ is assumed. The solid and dashed lines correspond to an integrated luminosity of 100 fb$^{-1}$ and 1000 fb$^{-1}$, respectively. The gray region is excluded by the kinematic condition $m_{H^+}+m_N>\sqrt{s}$ of single charged Higgs production at $\mu$TRISTAN.}
	\label{fig3}
\end{figure}

Based on the above analysis, we then explore the $5\sigma$ discovery region of the same-sign tetralepton signature at $\mu$TRISTAN. The results are shown in Figure \ref{fig3}. The free parameter set of this signature is $\{m_{H^+},m_N,y_{\mu N}\}$. In panel (a) of Figure \ref{fig3}, we fix the mass relation $m_{H^\pm}=2m_N$ to obtain the discovery limit of the Yukawa coupling $y_{\mu N}$. For the pair production dominant region $m_{H^+} \leq 1000$ GeV, 100 fb$^{-1}$ data could discover the same-sign tetralepton signature when $y_{\mu N} \gtrsim0.14$. Increasing the luminosity to 1000 fb$^{-1}$, the lower limit of $y_{\mu N}$ could further down to about 0.08 when $m_{H^+}\gtrsim400$ GeV. For light charged Higgs below 400 GeV and $y_{\mu N}\lesssim0.1$, the contribution from single production becomes also important as shown in Figure~\ref{fig2},  which enhances the discovery ability of this signal. Therefore, the lower limit of $y_{\mu N}$ clearly decreases as $m_{H^+}$ becomes smaller with 1000 fb$^{-1}$ and light charged Higgs below 400 GeV. For the single production dominant region $m_{H^+} \geq 1000$ GeV, the lower limit of $y_{\mu N}$ quickly increases as $m_{H^+}$ becomes larger. Therefore,  even with $y_{\mu N}=1$, we can only discover $m_{H^+}\sim 1130$ GeV via the same-sign tetralepton signal for 100~fb$^{-1}$ luminosity. The discovery upper limit on $m_{H^+}$ is extended to about 1250~GeV provided a luminosity of 1000 fb$^{-1}$.

In pale (b) of Figure \ref{fig3}, we show the $5\sigma$ discovery region in the $m_N-m_{H^+}$ plane with $y_{\mu N}=1$. For such a large Yukawa coupling, almost all the kinematically allowed region of pair production channel, i.e. $m_N<m_{H^+}\leq1000$ GeV, could be discovered by $\mu$TRISTAN. For the single production dominant region $m_{H^+}>1000$ GeV, the upper limit on $m_N$ sharply decreases as $m_{H^+}$ increases. With 100~fb$^{-1}$ data, we report that the upper limit on $m_{H^+}$  is approximately $1170$ GeV. Meanwhile, $\mu$TRISTAN could probe $m_{H^+}\lesssim1480$ GeV via the same-sign tetralepton signature with an integrated luminosity of 1000 fb$^{-1}$, which covers a large part of the kinematically allowed region.

\section{Conclusion}\label{SEC:CL}

The neutrinophilic two Higgs doublet model can naturally generate tiny neutrino mass at the TeV scale. This model introduces a new Higgs doublet $\Phi_\nu$ with $L_{\Phi_\nu}=-1$ and three heavy neutral leptons with $L_N=0$. The neutrinophilic scalar obtains a small VEV $v_\nu$ through the lepton number violation term $\mu^2(\Phi^\dag \Phi_\nu +\text{h.c.})$. Then the Yukawa interaction $y \overline{L} \tilde{\Phi}_\nu N$ generates the light neutrino via the seesaw mechanism. With small VEV $v_\nu$, the Yukawa coupling $y$ could be relatively large even with $m_N\sim \mathcal{O}$(TeV).

In this paper, we propose the novel same-sign tetralepton signature $4\mu^++4j$ at $\mu$TRISTAN in the neutrinophilic two Higgs doublet model. This signature is mediated by the muon-flavor Yukawa coupling $y_{\mu N}$. One contribution is from the pair production of charged Higgs as $\mu^+\mu^+\to H^+H^+ \to \mu^+ N + \mu^+ N\to 4\mu^++4j$, and the other one is from the single production of charged Higgs as $\mu^+ \mu^+ \to \mu^+ N H^+ \to \mu^+ N+ \mu^+ N \to 4\mu^++4j$. The former one is typically dominant when pair production is kinematically allowed $2 m_{H^+}<\sqrt{s}$ and $y_{\mu N}\gtrsim0.1$. Above the threshold, the single production channel becomes the only viable contribution. 

With quite a clean SM background, we study the inclusive same-sign tetralepton signal $4\mu^++\geq2j$ at $\mu$TRISTAN. For 100 (1000) fb$^{-1}$ luminosity, we could discover most parameter space within $y_{\mu N}\gtrsim0.14 (0.08)$ and $m_{H^+}\lesssim1130 (1250)$ GeV when $m_{H^+}=2m_N$. By fixing $y_{\mu N}=1$, we report that almost all the  kinematically allowed region of the pair production channel could be discovered with 100 fb$^{-1}$ data. The single production channel could further discover $m_{H^+}\lesssim1170$ GeV. Meanwhile, increasing the luminosity to 1000 fb$^{-1}$ could enlarge the discover region to $m_{H^+}\lesssim1480$ GeV.  Therefore, the same-sign tetralepton signature is promising at $\mu$TRISTAN, which provides an appealing pathway to test the TeV-scale origin of neutrino masses.

\section*{Acknowledgments}

This work is supported by the National Natural Science Foundation of China under Grant No. 12375074, State Key Laboratory of Dark Matter Physics, and University of Jinan Disciplinary Cross-Convergence Construction Project 2024 (XKJC-202404).


\end{document}